\documentclass[11pt,a4paper]{article}

\usepackage{amssymb}
\usepackage{amsmath}
\usepackage{graphicx}
\usepackage{enumerate}
\usepackage{eurosym}
\usepackage{hyperref}
\usepackage[ansinew]{inputenc}
\usepackage{verbatim}
\usepackage{color}

 \DeclareMathOperator*\esssup{ess\,sup}
\DeclareMathOperator*\essinf{ess\,inf}
\DeclareMathOperator*\argmin{arg\,min}
 \DeclareMathOperator\diag{diag}
\DeclareMathOperator\VaR{VaR}

\newcommand{\Real}{\mathbb{R}}

\newcommand{\E}{\mathbb{E}}

\newcommand{\specialcell}[1]{\ifmeasuring@#1\else\omit$\displaystyle#1$\ignorespaces\fi}

\title{{Portfolio Optimization under Shortfall Risk Constraint 
}}
\author{Oliver Janke\thanks{Humboldt-Universit\"{a}t zu Berlin, Department of Mathematics, Unter den Linden 6, 10099 Berlin, Germany. The authors thank Ulrich Horst, Julio Backhoff and Anna-Maria Hamm, as well as the anonymous referees for helpful suggestions and comments.} \thanks{{\em E-mail:} janke@math.hu-berlin.de; corresponding author.}
\and Qinghua Li\footnotemark[1]}

\begin{document}
\maketitle

\begin{abstract}
This paper solves a utility maximization problem under utility-based
shortfall risk constraint, by proposing an approach
using Lagrange multiplier and convex duality. 
Under mild conditions on the asymptotic elasticity of the utility
function and the loss function, we find an optimal wealth process
for the constrained problem and characterize the bi-dual relation
between the respective value functions of the constrained problem
and its dual. This approach applies to both complete and incomplete
markets. Moreover, the extension to more complicated cases is
illustrated by solving the problem with a consumption process added.
Finally, we give an example of utility and loss
functions in the Black-Scholes market where the solutions have
explicit forms.  
\end{abstract}


\fbox{\parbox{16.25cm}{
This is an Accepted Manuscript of an article published by Taylor \& Francis Group in {\em Optimization} on 19/04/2016, available online: \href{http://dx.doi.org/10.1080/02331934.2016.1173693}{http://www.tandfonline.com/10.1080/02331934.2016.1173693}
}}

\newtheorem{theorem}{Theorem}[section]
\newtheorem{proposition}[theorem]{Proposition}
\newtheorem{lemma}[theorem]{Lemma}
\newtheorem{definition}[theorem]{Definition}
\newtheorem{remark}[theorem]{Remark}
\newtheorem{example}[theorem]{Example}
\newtheorem{examples}[theorem]{Examples}
\newtheorem{corollary}[theorem]{Corollary}
\newtheorem{assumption}[theorem]{Assumption}
\newtheorem{property}[theorem]{Property}
\newtheorem{problem}[theorem]{Problem}
\newtheorem{def theorem}[theorem]{Definition and Theorem}
\newtheorem{definition lemma}[theorem]{Definition and Lemma}

\section{Introduction}

A portfolio manager strives to achieve two goals -- maximizing
profit and preventing risk. The former is formulated as maximizing
an expected utility from terminal wealth $X(T)$, where their
preference is modeled by a utility function $U$:
\begin{equation}\label{eqn1}
\max_{X} \, \E[U(X(T))].
\end{equation}
The latter is translated into a constraint on their risk measurement
$\rho$:
\begin{equation}\label{eqn2}
\rho(X(T)) \ \leq \ 0.
\end{equation}
The portfolio manager then solves a utility maximization problem under risk constraint.\\[-0.2cm]

The unconstrained version of utility maximization was first
introduced by Merton \cite{merton} who
solved the problem for power, logarithmic and exponential utility
functions where he found explicit solutions to the optimal trading strategy in case of two assets. 
Afterwards, Kramkov and Schachermayer \cite{kramkov schachermayer,
kramkov schachermayer2} developed the duality approach that solved
the problem in a general incomplete semimartingale model of the
financial market. Since Artzner et al.\ \cite{artzner et al}
mathematically defined measures of risk which were then developed
by, for example, F\"{o}llmer and Schied \cite{follmerschied}, portfolio
optimization under risk constraints has been an active topic of research.
Financial crises in the past decade raised even more alert to risks resulted from portfolio strategies.\\[-0.2cm]

On the other hand, also minimizing shortfall risk measure was widely studied in literature: Leibowitz and Henriksson \cite{leibowitz henriksson} introduced a confidence limit approach for for optimization problems with a shortfall constraint. Rockafellar and Uryasev \cite{rockafellar uryasev} showed that Conditional Value at Risk can be minimized by using linear programming and nonsmooth optimization techniques for a class of problems while calculating the Value at Risk. 
Moreover, an overview about important properties of the shortfall as a measure of risk as well as its comparison to other risk measures can be found in the works of Acerbi and Tasche \cite{acerbi tasche} and of Bertsimas et al.\ \cite{bertsimas et al}. 
Combining the results of \cite{follmerschied}, \cite{rockafellar uryasev} and others, Goldberg et al.\ \cite{goldberg et al} considered a problem of expected shortfall optimization. In contrast to previous works, they compared minimum expected shortfall to minimum variance portfolios instead of considering forecasting mean return and showed that downside risk optimization with the use of factor-based extreme risk is a realistic alternative to variance minimization. A recent paper by Bin \cite{bin} studied a portfolio optimization problem with a investor's stop strategy and derived a new Condional Value at Risk equation which leads to better solutions than the traditional method. \\[-0.2cm]

This paper will solve the utility maximization problem \eqref{eqn1}
under the constraint \eqref{eqn2}, with $\rho$ being a utility-based
shortfall risk measure. Our approach develops the convex duality for
utility maximization introduced by Kramkov and Schachermayer
\cite{kramkov schachermayer}. Under mild assumptions on the utility
function and the loss function, we show that the Lagrange function
is a usual utility function whose asymptotic elasticity is less than
one. An unconstrained maximization problem where the utility is the
Lagrange function can then be solved via the duality approach.
Solution to the constrained problem is shown to be the one to the
unconstrained problem with a proper choice of the Lagrange
multiplier. We provide an optimal wealth process and the bi-dual
relation between the respective value functions of
the constrained problem and a dual problem.\\[-0.2cm]

In the Black-Scholes framework where the price processes of the assets follow geometric Brownian motions we will consider a complete market where the number of shares equals the number of uncertainties. In this case, we derive a simpler form of the optimal solution as in the general case of semimartingale processes for the prices. Moreover, we shall give an example where the explicit solution for the optimal trading strategy is derived. \\[-0.2cm]

To illustrate extensions of our approach to more complicated cases,
we solve the optimal investment and consumption problem with
constraint on the utility-based shortfall risk. The unconstrained
version was first formulated and solved by Karatzas et al.\
\cite{karatzas et al} where the two problems were first considered
separately and then composed. Karatzas and Zitkovic \cite{karatzas
zitkovic} used time-dependent utility functions and extended the
notion of the asymptotic elasticity to this case. Using convex
duality techniques, they solved the pure consumption as well as the
combined consumption and terminal wealth problem. We shall add the
risk constraint to their version.\\[-0.2cm]

Similar problems on portfolio optimization under risk constraints have been investigated by other researchers as well. 
For instance, Basak and Shapiro \cite{basak shapiro} as well as Gabih et al.\ \cite{gabih grecksch wunderlich, gabih grecksch wunderlich 2} considered such a problem in a complete Black-Scholes financial market but where the risk measure is not cash-invariant. Gundel and Weber \cite{gundel weber, gundel weber 2}, Gabih et al.\ \cite{gabih sass wunderlich}, and Rudloff et al.\ \cite{rudloff sass wunderlich} used utility-based shortfall risk as risk measure. \\

A BSDE approach was formulated for example by Moreno-Bromberg et al.\ \cite{moreno et al} and Horst et al.\ \cite{horst et al}. Backhoff and Silva \cite{backhoff silva} analyzed connections between the Pontryagin's principle and Lagrange multiplier techniques for solving utility maximization problems under constraints. 
Cuoco et al.\ \cite{cuoco et al} considered semi-dynamic risk contraints in a complete Black-Scholes market where investors make at each time use of their information and evaluate their risk by applying the static risk measure to the conditional distribution of the projected change in wealth. \\[-0.2cm]

Although other authors also used a Lagrange multiplier to connect the utility function with the loss function (e.g.\ cf.\ \cite{gundel weber, gundel weber 2, gabih, rudloff sass wunderlich}), our approach explicitly takes the asymptotic elasticity into account: we define the properties for the asymptotic elasticity of the utility and the loss function such that there exists a solution for the optimization problem and give the conjugate function for the value function of the constrained optimization problem. \\

Moreover, we show that this value function is again a utility function and calculate it as well as its asymptotic elasticity. 
In contrast to Gundel and Weber \cite{gundel weber, gundel weber 2} we only consider wealth processes of admissible trading strategies instead of terminal wealths and therefore do not need a second Lagrange multiplier for the budget constraint. 
Moreover, our results can be extended to other portfolio optimization problems, which we show for the optimal consumption problem. 
To the best of our knowledge, this was not solved yet. 
The main advantage of our approach is that it can be also used for other loss functions except from shortfall risk, as long as the connected function is again a utility function. 
Last, we give a concrete example in a complete Black-Scholes market and solve the optimal trading strategy explicitly. 
In contrast to Gabih \cite{gabih} or Gundel and Weber \cite{gundel weber, gundel weber 2} we use utility and loss functions which do not have the same form.\\[-0.2cm]

The remainder of this paper is organized into five sections.
Section \ref{sec problem formulation} defines the financial market, the utility function and the risk measure. Our approach is proposed in a typical setting of utility maximization and utility-based shortfall risk measure. Moreover, we introduce the methodology using Lagrange multiplier to obtain another problem with a new utility function in subsection \ref{subsec portfolio optimization under risk constraint}. 
The original optimization problem under risk constraint is solved by linking it with an auxiliary problem in the incomplete market in subsection \ref{subsec main theorem incomplete market}. Moreover as an extension, we add a consumption process to the model in subsection \ref{subsec consumption}. 
We consider a complete market in section \ref{sec complete market}, where we derive the optimal solution in subsection \ref{subsec main theorem complete market}, extend the optimization problem in the Black-Scholes market by solving the problem in subsection \ref{sec extensions} and give an concrete example in subsection \ref{sec examples},  where we derive for a special utility and loss function an explicit form of the optimal wealth process and the optimal trading strategy.
The paper ends with a conclusion in section \ref{sec conclusion}.

\section{Problem formulation} \label{sec problem formulation}

\subsection{The market}
Let $(\Omega, \mathcal{F}, (\mathcal{F}_t)_{0 \leq t \leq T}, P)$ be
a filtered probability space. The time horizon of the financial
market is the interval $[0,T]$, for some positive real number $T$.
The market consists of one risk-free bond $S^0$ and $m$ stocks
$\tilde{S}=(\tilde{S}^1,\ldots,\tilde{S}^m)'$. With a deterministic
interest rate $r: [0,T] \rightarrow \mathbb{R}, t \mapsto r_t$, the
bond is given by $S^0_t = \exp{\left\{\int_0^t r_s ds\right\}}
> 0$, for all $t \in [0,T]$. Furthermore, the discounted stock price
process $S:=(S^1, \ldots, S^m)'$ with $S^i:=\tilde{S}^i/S^0$, $i=1,
\ldots,m$, is an $m$-dimensional semimartingale with respect to $(P,
(\mathcal{F}_t)_{0 \leq t \leq T})$.\\[-0.2cm]

Let $x$ denote the positive and exogenously given {\em initial
capital} of the investor. Let the {\em trading strategy}
$\pi=(\pi^1, \ldots, \pi^m)'$ be a predictable, $S$-integrable
process, where $\pi^i_t$, $i=1, \ldots, d$, denotes the number of
asset $i$ held in the portfolio at time $t$. The {\em portfolio} is
defined as the pair $(x,\pi)$. The associated {\em wealth process}
is denoted as $X^{\pi,x}$. The leftover wealth $X^{\pi,x} -
\sum_{i=1}^d \pi^i$ is invested in the risk-free bond. Our portfolio
$(x,\pi)$ is {\em self-financing}, in the sense that there will be
no exogenous cash-flow like credits or consumption. Consequently,
the wealth process is given by
\begin{eqnarray} \label{eq wealth process}
X^{\pi,x}(t) & = & x + \int_0^t {\pi'_u} \ dS_u , \quad 0 \leq t
\leq T.
\end{eqnarray}
The set of all such admissible trading strategies $\pi$ is denoted
as $\Pi$. The set of all nonnegative wealth processes with initial
capital $x$ is defined as
$$ \mathcal{X}(x) \ := \ \left\{ \ X^{\pi,x}\geq 0 \ \left| \ \pi \in \Pi \ \right. \right\} . $$
When there is no confusion, we simply write $X$ for $X^{x,\pi}$.

\begin{definition}
The set $\mathcal{Q}$ of {\em equivalent local martingale measures}, with
respect to the probability measure $P$ and the wealth process set
$\mathcal{X}(1)$, is the collection of all probability measures $Q$
which satisfy \\
(i) $P$ and $Q$ are equivalent ($Q \sim P$); \\ 
(ii) any $X \in \mathcal{X}(1)$ is a local martingale under $Q$.
\end{definition}

If the price process $S$ is locally bounded, then it is a local
martingale under any equivalent local martingale measure $Q$ on
$[0,T]$. Moreover, we denote by $D(\mathcal{Q})$ the set of all
Radon-Nikodym derivatives $dQ/dP$ for any probability measure $Q \in
\mathcal{Q}$ with respect to $P$.

\begin{assumption} \label{ass nonempty q} {\em
We assume throughout the paper that $ \mathcal{Q} \neq \ \emptyset$.
}
\end{assumption}

Economically, the existence of an equivalent local martingale
measure is equivalent to the absence of arbitrage in the following
sense.

\begin{def theorem}(\cite{delbaen schachermayer}, Corollary 1.2)
Let $S$ be a locally bounded real-valued semimartingale. There is an equivalent local martingale measure for $S$ if and only if $S$ satisfies
{\em No Free Lunch with Vanishing Risk}, i.e., there is no sequence $(f_n)_{n \geq 0}$ of final payoffs of admissible integrands, $f_n = \int \pi_n dS$, such that the negative parts $f_n^-$ tend to 0 uniformly and such that $f_n$ tends almost surely to a $[0,+\infty]$-valued function $f_0$ satisfying $P(f_0 > 0) > 0$.
\end{def theorem}

The market is complete when the equivalent local martingale measure
is unique (cf.\ \cite{kramkov schachermayer}). Kardaras and Platen
\cite{kardaras platen} pointed out that the assumption of an
arbitrage-free market implies that the price process has to be a
semimartingale. Therefore, our semimartingale assumption for $S$ is
necessary. However, the contrary is not true, cf.\ \cite{kardaras
platen}. Hence we need Assumption \ref{ass nonempty q}.

\subsection{Utility functions} Now, let us introduce the exogenous
time and state independent utility function of the investor who
receives a certain cash amount from each investment strategy.
Intuitively, the utility function $U$ compares the satisfactory of
the investor brought by different cash amounts. Rigorously, a
utility function $U$  is defined in the definition below.

\begin{definition}[Utility function] \label{def utility function}
Let a function $U:(0,+\infty) \rightarrow \Real \cup \{-\infty\}$,
$x\mapsto U(x)$ be given. $U$ is called a {\em utility function}, if
it is strictly increasing, strictly concave, continuously differentiable and if it satisfies the \textbf{Inada} conditions
$$ U'(+\infty):= \lim_{x \rightarrow \infty} U'(x) = 0 \quad \mbox{ and } \quad U'(0) := \lim_{x \searrow 0} U'(x) = +\infty. $$
\end{definition}

Moreover, the inverse function of the first order derivative of $U$ is denoted by $I:=(U')^{-1}$. \\[-0.2cm]

The Legendre transform $V$ of $-U(-x)$ is very useful in solving a
utility maximization problem and calculating the optimal terminal
wealth (cf.\ \cite{karatzas et al 2, rockafellar}). It is given by
\begin{equation} \label{eq legendre transform}
V(y) \ := \ \sup_{x > 0} \{U(x)-xy\} \ = \ U(I(y)) - yI(y), \quad 0
< y < +\infty.
\end{equation}

The bi-dual relation is given by
\begin{equation}
U(x) \ = \ \inf_{y > 0} \, \{V(y) + xy \}, \quad x > 0.
\end{equation}

The following result describes the asymptotic properties of the
Legendre transform $V$. The proof can be found for example in
\cite[Lemma 4.2]{karatzas et al 2}.

\begin{property}\label{property dual V}
Suppose $U$ is a utility function defined in Definition \ref{def
utility function}, then the function $V$ defined in \eqref{eq
legendre transform} is continuously differentiable, decreasing,
strictly convex and satisfies
$$ V'(+\infty)\ := \ \lim_{y \rightarrow \infty} V'(y) \ = \ 0 \quad \mbox{ and } \quad V'(0) \ := \ \lim_{y \searrow 0} V'(y) \ = \ -\infty. $$
Moreover, it holds that
$$ V(0) \ := \ \lim_{y \searrow 0} V(y) \ = \ U(+\infty) \quad \mbox{ and } \quad V(+\infty) \ := \ \lim_{y \rightarrow \infty} V(y) \ = \ U(0). $$
The inverse function $I$ of the first order derivative of $U$
satisfies $I := (U')^{-1} = -V'$.
\end{property}

\subsection{Risk measures}
For a given amount of initial capital, the agent's trading strategy
is restricted by their risk preference. Therefore, we assume that
the agent is risk averse and that the risk, measured by a specific
function, is bounded from above. A risk measure is defined by its
properties. Giesecke et al.\ \cite{giesecke et al} pointed out that a
good risk measure should quantify risk on a monetary scale, detect
the risk of extreme loss events and encourage diversification of
portfolio choice.

\begin{definition}[Convex risk measure](\cite{giesecke et al}, Definition 2.3) \label{def static convex risk measure}
Let $\mathcal{X}$ be some vector space of integrable random
variables. The functional $\rho: \mathcal{X} \rightarrow \mathbb{R}$
is called a {\em (monetary) convex risk measure} if the properties
(a,b,c) hold true for any $X_1, X_2 \in \mathcal{X}$. \\
(a) {\em Convexity:} $\rho(\lambda X_1 + (1-\lambda) X_2) \leq \lambda \rho(X_1) + (1-\lambda) \rho(X_2)$, for any $\lambda \in [0,1]$. \\
(b) {\em Monotonicity:} $X_1 \leq X_2$ implies $\rho(X_2) \leq \rho(X_1)$. \\
(c) {\em Translation invariance:} $\rho(X_1 + m) = \rho(X_1) - m$, for any $m \in \mathbb{R}$.
\end{definition}

By property translation invariance (c), the value $\rho(X)$, $X \in
\mathcal{X}$, can be interpreted as the value which an agent must
add to their risky asset $X$ to eliminate the risk. To wit,
$$\rho(X + \rho(X)) \ \stackrel{(c)}{=} \ \rho(X) - \rho(X) \ = \ 0. $$
The interpretation of the property convexity (a) is that an agent
can minimize the risk by diversifying their portfolio.
Monotonicity (b) means that the risk decreases if the payoff profile
is increased.
There also exists a dynamic version of risk measures, which is
defined by, for example, F\"{o}llmer and Schied \cite{follmerschied}. \\[-0.2cm]

A very famous risk measure often used in the financial industry is
the {\em Value at Risk} (VaR). For a financial position $X \in
\mathcal{X}$, it is defined as the smallest value $m \in \mathbb{R}$
which has to be added to $X$ such that the probability of a loss
does not exceed a given level $\alpha \in (0,1)$. Mathematically,
VaR is expressed as (cf.\ \cite{giesecke et al})
\begin{eqnarray*}
\VaR_{\alpha}(X) & := & \inf \{ m \in \mathbb{R} \ \left| \ P(X + m
< 0) \, \leq \, \alpha \right.\}.
\end{eqnarray*}

Although it is often used in banks and insurance companies, VaR has
some disadvantages. First, it does not take into account the size of
losses exceeding the VaR. Second, the convex property in Definition
\ref{def static convex risk measure} (a) does not hold for VaR in
general, so it does not encourage diversification. Since we focus on
convex risk measures, our approach does not cover the VaR case.
Nevertheless, it was solved for a Black-Scholes market by Basak and Shapiro \cite{basak
shapiro}. \\
To avoid its disadvantages, VaR can be modified into {\em Average
Value at Risk} (AVaR), cf.\ \cite{giesecke et al}. 


In this article, we refer to a special risk measure which was first introduced by Föllmer and Schied \cite{follmerschied} defined through a loss function.

\begin{definition}[Loss function]\label{def loss function}
A function $L:(-\infty,0) \rightarrow \mathbb{R}$ is called a {\em
loss function}, if it is a strictly increasing and strictly convex
function  satisfying the following properties.\\
(i) $L$ is continuous differentiable on $(-\infty,0)$. \\
(ii) $\lim_{x \rightarrow 0} L'(x)>-\infty$ and $\lim_{x \rightarrow -\infty} L'(x)=0$.
\end{definition}

Through this loss function, we can define a {\em utility-based
shortfall risk measure} as the smallest capital amount $m \in
\mathbb{R}$ which has to be added to the position $X$, such that the
expected loss function of it stays below some given value $x_1$.

\begin{definition lemma}[Utility-based shortfall risk]\label{def short fall} (\cite{follmerschied}, p.\ 8-9)
A risk measure $\rho^L$ is called {\em utility-based shortfall risk}, if there exists a loss function $L$ defined according to Definition \ref{def loss function}, such that $\rho^L$ can be written in the form of
$$ \rho^L(X) \ = \ \inf \left\{ m \in \mathbb{R} \ \left| \ \mathbb{E}[L(-X - m)] \, \leq \, x_1 \right. \right\}. $$
Then requiring that $\rho^{L}(X)\leq 0$ is equivalent to requiring that $\mathbb{E}[L(-X)] \leq x_1$.
\end{definition lemma}

{\em Proof.} \\
{\em if-part:} Let $\rho^{L}(X)\leq 0$. Then it holds due to the strict increase of $L$ that
$$ x_1 \ \geq \ \E[L(-X-\rho^{L}(X))] \ = \ \E[L(\underbrace{-(X+\rho^{L}(X))}_{\geq -X})] \ \geq \ \E[L(-X)]. $$
{\em only if-part:} Let $\mathbb{E}[L(-X)] \leq x_1$. Then
$\rho^{L}(X)=0$ satisfies $\mathbb{E}[L(-X-\rho^{L}(X))] \leq x_1$,
so $\rho^{L}(X) = \inf \left\{ m \in \mathbb{R} \mid \mathbb{E}[L(-X - m)] \leq x_1 \right\} \leq 0$.
\hfill{$\Box$}

\begin{example}[Entropic risk measure] \label{ex entropic risk measure}
If we consider a function of the exponential form $L(x)=\exp\{\gamma
x\}$, where $\gamma > 0$ represents the risk aversion of the
investor, then all properties in Definition \ref{def loss function}
are satisfied, so $L$ is a loss function. The associated risk
measure $e_{\gamma}$, given by
\begin{eqnarray} \label{def entropic risk}
e_{\gamma}(X) & := & \frac{1}{\gamma} (\ln \mathbb{E} [\exp\{-\gamma
X\}] - \ln x_1),
\end{eqnarray}
is called the {\em entropic risk measure} (cf.\ \cite{rudloff sass wunderlich}).
\end{example}

{\em Remark.} 
The result that an elicitable\footnote{A law-invariant risk measure $\rho$ is called {\em elicitable} if there exists a scoring function $S:\mathbb{R}^2 \to \mathbb{R}$ such that $\rho(F) = \argmin_{x \in \mathbb{R}} S(x,y) dF(y)$ for any probability measure $F$ on $\mathbb{R}$, cf.\ \cite{bellini bignozzi}.} risk measure $\rho$ is a shortfall was first shown by Weber \cite{weber} under some regulatory conditions and without assuming that the loss function $L$ is convex. 
Bellini and Bignozzi \cite{bellini bignozzi} proved the statement under weak assumptions on the scoring function. 
They showed that if $\rho$ is elicitable and convex, then $L$ is also convex. 
Moreover, if $\rho$ is additionally positive homogenous, i.e., $\rho(\lambda X)=\lambda \rho(X)$ for $X \in \mathcal{X}$ and $\lambda \geq 0$, then $L$ is of the form $L(x)=L_0 + \alpha x^+ -\beta x^-$ for $\alpha \geq \beta > 0$. 
Such risk measures are called {\em expectiles}, see \cite{newey powell}. 
The results in \cite{weber} were also extended by Delbaen et al.\ \cite{delbaen et al} by showing that convex law-invariant risk measures correspond to generalized shortfalls, in which the loss function can also take the value infinity. 


\subsection{Portfolio optimization under risk constraint} \label{subsec portfolio optimization under risk constraint}

Let $x>0$ be the initial capital. The utility function $U$ and the
loss function $L$ are given. This paper aims at solving the
following portfolio optimization problem under utility-based
shortfall risk constraint.

\begin{problem}\label{problem1}
Find an optimal wealth process $\tilde{X}$ that achieves the maximum expected utility
\begin{equation} \label{utility max}
u(x) \ := \ \sup \limits_{X \in \mathcal{A}(x)} \mathbb{E} \left[U(X(T)) \right].
\end{equation}

For a given benchmark $x_1$, the set
\begin{equation}
\mathcal{A}(x)\ := \ \left\{ X \in \mathcal{X}(x) \ \left| \ \mathbb{E}[L(-X(T))] \leq x_1 \right.\right\}
\end{equation}
is the set of admissible wealth processes that satisfy the
constraint on the utility-based shortfall risk. The function
$u(\cdot)$ is called the ``value function" of this optimization
problem.
\end{problem}

To exclude trivial cases we assume throughout the paper that
\begin{equation} \label{ass u finite}
\begin{split}
& \sup \limits_{X \in \mathcal{X}(x)} \mathbb{E} \left[ U(X(T))\right]  \ < \  +\infty, \quad \text{for some } x > 0; \\
& \inf\limits_{X \in \mathcal{X}(x)} \mathbb{E} \left[ L(-X(T)) \right] \ > \ -\infty, \quad \text{for all } x > 0.
\end{split}
\end{equation}

It is easy to imagine that there will not be a solution to this optimization problem for all $x_1$.
On the one hand, the restriction could be too strong that there is no trading strategy such that the corresponding terminal wealth $X(T)$ for $X \in \mathcal{X}(x)$ satisfies the risk constraint.
On the other hand, the restriction could also be too weak such that the risk constraint is not binding. To be more precise, let us define
\begin{eqnarray*}
r_{\min} & := & \inf_{X \in \mathcal{X}(x)} \, \left\{ \E[L(-X(T))] \right\}  \qquad \text{and} \\
r_{\max} & := & \sup_{X \in \mathcal{X}(x)} \, \left\{ \E[L(-X(T))]\ \left| \ \E[U(X(T))] \geq \E[U(X^{\#}(T))] \mbox{ for any } X^{\#} \in \mathcal{X}(x) \right. \right\}.
\end{eqnarray*}

In special cases, we can explicitly express $r_{\min}$ and $r_{\max}$ (cf.\ \cite{gundel weber 2}, Lemma 6.1.). \\

From now on, for a given $x > 0$ we choose $x_1$ such that $r_{\min} \leq x_1 \leq r_{\max}$. \\

Because this is an optimization problem under constraints, we shall
reformulate it by introducing a Lagrange multiplier $\lambda \geq 0$
(cf.\ \cite{rockafellar}). Let us define this new function
$W_\lambda:(0, +\infty) \to \Real \cup \{-\infty\}$ by
\begin{equation} \label{def W}
W_\lambda(X) \ := \ U(X) - \lambda L(-X), \quad \lambda > 0.
\end{equation}
By the definitions of $U$ and $L$, we have the following properties of $W_\lambda$.

\begin{proposition} \label{prop properties w}
Let $W_\lambda$ be a function as defined in \eqref{def W}. Then \\
(a) $W_\lambda$ is strictly increasing, strictly concave and continuously differentiable on
$(0,+\infty)$;\\
(b) $W_\lambda$ satisfies the Inada conditions
$$ W_\lambda'(+\infty) \ := \ \lim_{x \rightarrow \infty} W_\lambda'(x) \ = \ 0 \quad \mbox{ and } \quad W_\lambda'(0) \ := \ \lim_{x \searrow 0} W_\lambda'(x) \ = \ +\infty. $$
\end{proposition}

{\em Proof.}
\begin{enumerate}
\item[(a)] If $\lambda=0$, the proof it obvious. Now assume that $\lambda > 0$. Because $L(x)$ is strictly increasing and strictly convex in $x$, $-\lambda L(x)$ is strictly decreasing and strictly convex and $-\lambda L(-x)$ is strictly increasing and strictly concave in $x$ for any $\lambda > 0$. Moreover, $-L(-x)$ is continuously differentiable on $(0,+\infty)$, because $L$ is continuously differentiable on $(-\infty,0)$. Therefore, the sum $U(x) - \lambda L(-x)$ is a strictly increasing and concave function, which is continuously differentiable on $(0,+\infty)$ for any $\lambda > 0$.
\item[(b)] Due to part (a) and the assumptions on $U$ and $L$, it holds for any $\lambda>0$ that
\begin{align*}
\lim_{x \rightarrow \infty} W_\lambda'(x) = \lim_{x \rightarrow \infty} (U'(x) + \lambda L'(-x)) = 0, \
\lim_{x \searrow 0} W_\lambda'(x) = \lim_{x \searrow 0} (U'(x) + \lambda L'(-x)) = +\infty.  \quad \Box
\end{align*}
\end{enumerate}

$W_\lambda$ has the same properties as a usual utility function $U$ defined in Definition \ref{def utility function}.
Therefore, we can use Property \ref{property dual V} and introduce the conjugate function $Z_\lambda$ of $W_\lambda$ by
\begin{equation} \label{def Z}
 Z_\lambda(y) \ := \ \sup_{x>0}\, \{W_\lambda(x) - xy \} , \quad y > 0,
\end{equation}
which is the Legendre transform of $-W_\lambda(-x)$. The bi-dual relation is given by
$$ W_\lambda(x) \ = \ \inf_{y > 0} \, \{Z_\lambda(y) + xy \}, \quad x > 0. $$
According to Property \ref{property dual V}, $Z_\lambda$ is
continuously differentiable, decreasing, strictly convex and
satisfies
\begin{equation}
 Z_{\lambda}(0) \ = \ W_{\lambda}(+\infty) \quad \mbox{ and } \quad Z_{\lambda}(+\infty) \ = \ W_{\lambda}(0).
\end{equation}
Moreover, by the properties of $W_\lambda$, the inverse function
$H_\lambda$ of its first order derivative exists and satisfies
\begin{equation}\label{def H}
 H_\lambda \ := \ (W_\lambda')^{-1} \ = \ -Z_\lambda'.
\end{equation}
In sections \ref{sec incomplete} and \ref{sec complete market}, we shall
show that the optimal wealth process to Problem \ref{problem1} is
the one to the following unconstrained utility maximization problem
with a proper choice of the Lagrange multiplier $\lambda$.

\begin{problem}\label{problem1.1}
Let $W_\lambda$ play the role of a utility function. Find an optimal
wealth process $\tilde{X}_\lambda$ that achieves the maximum
expected utility
\begin{eqnarray} \label{def primal value function} w_\lambda(x) & :=
& \sup_{X \in \mathcal{X}(x)} \, \E\left[W_{\lambda}(X(T))\right].
\end{eqnarray}
\end{problem}

\begin{lemma} \label{lemma w finite}
Let \eqref{ass u finite} and \eqref{def W} hold true. Then there
exists an $x>0$, such that
$$ w_{\lambda}(x) \ < \ + \infty, \quad \text{for all }\lambda \geq 0. $$
\end{lemma}

{\em Proof.} Let $x > 0$ be such that $\sup_{X \in
\mathcal{X}(x)}E\left[U(X(T))\right] <+\infty$ as in \eqref{ass u
finite}. By equations \eqref{ass u finite} and \eqref{def W}, we
have for the value function that
 $$ w_{\lambda}(x) \ = \ \sup_{X \in \mathcal{X}(x)} \,
 \E\left[W_{\lambda}(X(T))\right] \
 \leq \ \sup_{X \in \mathcal{X}(x)} \, \E\left[U(X(T))\right]
 - \lambda \inf_{X \in \mathcal{X}(x)} \, \E\left[L(-X(T))\right]
 \ < \ + \infty. \quad {\Box} $$
$ $
\begin{lemma}\label{lemma property Z}
The functions $Z_\lambda$ and $H_\lambda$ defined in \eqref{def Z}
and \eqref{def H} have the following properties.
\begin{enumerate}
\item[(i)] Fixing any $y\in (0,\infty)$, the quantity $H_\lambda(y)$ is the unique solution to the equation
$$ U'(x) + \lambda L'(-x) \ = \ y $$
over the interval $x\in(0,\infty)$.
\item[(ii)] Assume that $L$ is positive-valued (resp.\ non negative-valued) and let $V$ be the Legendre transform defined in \eqref{eq legendre transform}, then the comparison
\begin{equation}
Z_\lambda(y) \ < \ V(y) \qquad (\text{resp.\ } \ Z_\lambda(y) \ \leq \ V(y))
\end{equation}
holds for all $y \in (0,+\infty)$.
\end{enumerate}
\end{lemma}

{\em Proof.}
\begin{enumerate}
\item[(i)] It follows by the definition of $H_\lambda$ in \eqref{def H}, that $H_\lambda(y)$ solves the equation $W'_{\lambda}(x)=y$. By the definition of $W_{\lambda}$ (cf. \eqref{def W}) it follows that $H_{\lambda}$ also solves $U'(x) + \lambda L'(-x) = y$. The uniqueness follows from the strict monotonicity of $W_\lambda$, cf.\ Proposition \ref{prop properties w}.
\item[(ii)] Since $W_{\lambda}$ is a utility function, we can use Property \ref{property dual V} to derive the conjugate function $Z_{\lambda}$ of $W_{\lambda}$. By the equations (\ref{def W}) and (\ref{def Z}), we know that
\begin{equation}\label{eqn Z 2}
Z_\lambda(y) \ = \ \sup_{x>0}\, \{W_{\lambda}(x) - xy \} \nonumber \ = \ \sup_{x>0}\, \{U(x)- \lambda L(-x) - xy \} , \quad y > 0.
\end{equation}
On the other hand, $V$ is the conjugate function of $U$, so it holds that
$$ V(y) \ = \ \sup_{x>0}\, \{U(x) - xy \}, \quad y > 0. $$
Because $L$ is positive (resp.\ non negative) by assumption, the identity
\begin{equation} \label{eqn Z 3}
U(x)- \lambda L(-x) - xy \ < \ U(x) - xy \qquad (\text{resp. } U(x)- \lambda L(-x) - xy \ < \ U(x) - xy)
\end{equation}
holds for all $x > 0$, $y > 0$ and $\lambda >0$.
The expressions (\ref{eq legendre transform}), (\ref{eqn Z 2}) and (\ref{eqn Z 3}) imply that $Z_\lambda(y) \leq V(y)$.
If $L$ is strictly positive, the strict inequality $Z_\lambda(y) < V(y)$ holds, because the suprema in the equations (\ref{eq legendre transform}) and (\ref{eqn Z 2}) are attained.
\hfill{$\Box$}
\end{enumerate}

\section{Solution in incomplete market}\label{sec incomplete}

In the case of an incomplete market, i.e., $|\mathcal{Q}|>1$, and
following the ideas of \cite{delbaen schachermayer, kramkov
schachermayer}, we have to dualize Problem \ref{problem1.1}. Thereby
we define
$$ \mathcal{Y}(y) \ := \ \{Y \geq 0 \, | \, Y(0)=y, \, XY=(X_tY_t)_{0\leq t\leq T} \mbox{ is a supermartingale for all } X \in \mathcal{X}(1)\} $$
as the set of nonnegative semimartingales $Y$ with $Y(0)=y$ and such
that the process $XY$ is a supermartingale for any $X \in
\mathcal{X}(1)$. In particular, due to the fact that $X \equiv 1$
belongs to $\mathcal{X}(1)$, any $Y \in \mathcal{Y}(y)$ is a
supermartingale. We note that the density process $dQ/dP$ of all
equivalent martingale measures $Q \in \mathcal{Q}$ also belongs to
$\mathcal{Y}(1)$. By Assumption
\ref{ass nonempty q}, the existence of at least one element of $\mathcal{Q}$ implies that $\mathcal{Y}$ is nonempty. \\[-0.2cm]

Let us now define the value function of the dual problem by
\begin{equation} \label{def dual value function}
z_\lambda(y) \ = \ \inf_{Y \in \mathcal{Y}(y)} \ \E \left[Z_\lambda(Y_T)\right].
\end{equation}

\subsection{Conditions on the asymptotic elasticity}

As pointed out by \cite{kramkov schachermayer}, a sufficient
condition for the existence of an optimal solution to an
unconstrained utility maximization problem in an incomplete market
is that the asymptotic elasticity of the utility function is less
than one. Economically, the elasticity $e(x)$ describes the relation
between the relative change of the output and the relative change of
the input. It is defined as
$$ e(x) \ := \ \frac{x U'(x)}{U(x)} \
= \ \lim\limits_{\Delta x\rightarrow 0} \frac{\frac{\Delta
U}{U}}{\frac{\Delta x}{x}} \ = \ \lim\limits_{\Delta x\rightarrow
0}\frac{x \frac{\Delta U}{\Delta x}}{U}. $$

The asymptotic elasticity is the upper limit of the elasticity when
$x$ tends to infinity.

\begin{definition}[Asymptotic elasticity] Let a utility function $U$ as defined in Definition \ref{def utility function} be given. The {\em asymptotic elasticity} $AE(U)$ of $U$ is defined by
$$ AE(U) \ := \ \limsup_{x \rightarrow +\infty} \frac{xU'(x)}{U(x)}. $$
Analogously, the asymptotic elasticity $AE_-(L)$ towards negative infinity for a given loss function $L$ as defined in Definition \ref{def loss function} is given by
$$ AE_-(L) \ := \ \limsup_{x \rightarrow -\infty} \frac{xL'(x)}{L(x)} \ = \ \limsup_{x \rightarrow +\infty} \frac{-xL'(-x)}{L(-x)}. $$
\end{definition}

We have a nice property about the range of the asymptotic elasticity depending on $U(+\infty)$.

\begin{lemma}(\cite{kramkov schachermayer}, Lemma 6.1) \label{lemma domain ae(u)}
For a strictly concave, increasing and real-valued function $U$ the
asymptotic elasticity $AE(U)$ is well-defined. The range of $AE(U)$
differs according to $U(+\infty):=\lim_{x \to \infty} U(x)$. \\
(i) If $U(\infty)=+\infty$, it holds that $AE(U) \in [0,1]$. \\
(ii) If $U(\infty) \in (0,+\infty)$, it holds that $AE(U) = 0$. \\
(iii) If $U(\infty) \in (-\infty,0]$, it holds that $AE(U) \in [-\infty, 0]$. 
\end{lemma}

Moreover, the asymptotic utility does not change for affine transformations of the utility function. This result was established in \cite{kramkov schachermayer} and the proof is easily verified. 

\begin{lemma}
Let $U$ be a utility function as defined in Definition \ref{def utility function} and let
its affine transformation function be given by $\tilde{U}(x)=c_1 + c_2 U(x)$,
where $c_1, c_2 \in \Real$ and $c_2 > 0$. If $U(+\infty)>0$ and $\tilde{U}(+\infty)>0$, then it holds that $AE(U) = AE(\tilde{U}) \in [0,1]$. 
\end{lemma}


For our constraint problem, it means that the asymptotic elasticity
of the function $W_\lambda$ must be less than one. The next lemma
tells us the conditions on $U$ and $L$ under which this will hold.

\begin{lemma} \label{lemma ae of w}
For the asymptotic elasticity $AE(W_\lambda)$ of $
W_\lambda(x):=U(x)-\lambda L(-x)$, $\lambda \geq 0$, we have the
following results. \\
(a) If $\lim_{x \to \infty} W_{\lambda} < +\infty$, equivalently if $U(+\infty)<+\infty$ and $L(-\infty)>-\infty$,
then $AE(W_\lambda) < 1$. \\
(b) For $\lim_{x \to \infty} W_{\lambda} = +\infty$ we have $AE(W_\lambda) < 1$ if one of
the following three cases holds true.
\begin{itemize}
\item $U(+\infty)=+\infty$, $L(-\infty)>-\infty$ and $AE(U) < 1$;
\item $U(+\infty)=+\infty$, $L(-\infty)=-\infty$, $AE(U) < 1$ and $AE_-(L) < 1$;
\item $U(+\infty)<+\infty$, $L(-\infty)=-\infty$ and $AE_-(L) < 1$.
\end{itemize}
\end{lemma}

{\em Proof.}
\begin{enumerate}
\item[(a)] It holds due to Lemma \ref{lemma domain ae(u)}.
\item[(b)] Due to the properties that $U'(x) \geq 0$, $\lim_{x \to \infty} U'(x)=0$,
$L'(x) \geq 0$ and $\lim_{x \to -\infty} L'(x)=0$, we can distinguish three cases. \\
{\em Case 1}: $U(+\infty)=+\infty$ and $L(-\infty)>-\infty$. Then it holds for any $\lambda > 0$ that
\begin{eqnarray*}
AE(W_\lambda) & = & \limsup_{x \to \infty} \frac{xW'_\lambda(x)}{W_\lambda(x)} \ = \ \limsup_{x \to \infty} \frac{x(U'(x) + \lambda L'(-x))}{U(x) - \lambda L(-x)} \\
& \leq & \limsup_{x \to \infty} \frac{xU'(x)}{U(x) - \lambda L(-x)} + \limsup_{x \to \infty} \frac{x\lambda L'(-x)}{U(x) - \lambda L(-x)} \\
& \leq & \limsup_{x \to \infty} \frac{xU'(x)}{U(x)} + \underbrace{\limsup_{x \to \infty} \frac{x\lambda L'(-x)}{U(x)}}_{=0} \ \leq \ AE(U).
\end{eqnarray*}
{\em Case 2}: $U(+\infty)=+\infty$ and $L(-\infty)=-\infty$. 
For any $\varepsilon \in (0,1-\max\{AE(U),AE_-(L)\})$, there exists $\bar{x} \in (0,+\infty)$ such that for all $x > \bar{x}$ it holds that
$$ \left\{ \begin{array} {rcl} -L(-x) \ > \ 0 & ; & U(x) \ > \ 0; \\
\frac{xU'(x)}{U(x)} & < & AE(U) + \varepsilon; \\
\frac{-xL'(-x)}{L(-x)} & < & AE_-(L) + \varepsilon. \end{array}
\right. $$ With this, it follows for all $x > \bar{x}$ that
$$ \left\{ \begin{array} {rcl}
xU'(x) & < & (\max\{AE(U),AE_-(L)\} + \varepsilon) U(x); \\
xL'(-x) & < & -(\max\{AE(U),AE_-(L)\} + \varepsilon) L(-x). \end{array} \right. $$
Moreover, we get for all $x > \bar{x}$ that
$$ \frac{xW'_\lambda(x)}{W_\lambda(x)} \ = \ \frac{xU'(x) + \lambda x L'(-x)}{U(x) - \lambda L(-x)} \ < \ \max\{AE(U), AE_-(L)\} + \varepsilon \ < \ 1, $$
and by the definition of $\limsup$ it holds that
$$ AE(W_\lambda) \ = \ \limsup_{x \to \infty} \frac{xW'_\lambda(x)}{W_\lambda(x)} \ \leq \ \max\{AE(U), AE_-(L)\} + \varepsilon \ < \ 1. $$
{\em Case 3}: $U(+\infty)<+\infty$ and $L(-\infty)=-\infty$. Then it holds for any $\lambda > 0$ that
\begin{eqnarray*}
AE(W_\lambda) & \leq & \limsup_{x \to \infty} \frac{xU'(x)}{U(x) - \lambda L(-x)} + \limsup_{x \to \infty} \frac{x\lambda L'(-x)}{U(x) - \lambda L(-x)} \\
& \leq & \underbrace{\limsup_{x \to \infty} \frac{xU'(x)}{-L(x)}}_{=0} + \limsup_{x \to \infty} \frac{x\lambda L'(-x)}{-\lambda L(-x)} \ \leq \  AE_-(L). \qquad \qquad \qquad {\Box}
\end{eqnarray*}
\end{enumerate}

Let us consider a special loss functions as a example for the asymptotic elasticity of $W_{\lambda}$.

\begin{example} 
If the loss function is of exponential form, i.e.\ $L(x)=e^{\gamma x}$, $\gamma > 0$, then for any utility function $U$ with $AE(U) < 1$ it holds that
\begin{eqnarray*}
\lefteqn{\limsup_{x \to \infty} \frac{x(U(x) - \lambda L(-x))'}{U(x) - \lambda L(-x)} \ = \ \limsup_{x \to \infty} \frac{x(U'(x) + \lambda \gamma e^{-\gamma x})}{U(x) - \lambda e^{\gamma x}}} \\
& \leq & \limsup_{x \to \infty} \frac{xU'(x)}{U(x) - \lambda e^{-\gamma x}} + \limsup_{x \to \infty} \frac{\lambda e^{-\gamma x}}{U(x) - \lambda e^{\gamma x}} \ \stackrel{L(-\infty)=0}{\leq} \ \limsup_{x \to \infty} \frac{xU'(x)}{U(x)} \ < \ 1, \ \mbox{ for any } \lambda > 0.
\end{eqnarray*}
\end{example}

\subsection{Main theorem} \label{subsec main theorem incomplete market}

We solve the auxiliary Problem \ref{problem1.1} and derive a unique optimal solution for it. 
Moreover, we show that there exists $\lambda^{*} \geq 0$ such that the risk constraint is exactly satisfied. 
With this, we solve Problem \ref{problem1} by connecting the value functions $w{_{\lambda^*}}$ and $u$. 
We show that $u$ is also a utility function satisfying all the conditions of Definition \ref{def utility function} with asymptotic elasticity strictly smaller than 1.

\begin{theorem} \label{lemma incomplete market}
Let Assumption \ref{ass nonempty q}, \eqref{ass u finite} and
\eqref{def W} hold true. Let furthermore the asymptotic elasticity
of $W_\lambda$ be strictly less than one.
Then we have the following results.
\begin{enumerate}
\item[(i)] Let $\tilde{X}_{\lambda}$ be the optimal solution to Problem \ref{problem1.1} and
 $\lambda^{*} \geq 0$ be such that
$\E[L(-\tilde{X}_{\lambda^*}(T))] = x_1$. Let $y=u'(x)$ and
$\tilde{Y}_{\lambda^*} \in \mathcal{Y}(y)$ be the unique optimal
solution to \eqref{def dual value function} with
$\lambda=\lambda^*$. The unique optimal solution $\tilde{X} \in
\mathcal{A}(x)$ to Problem \ref{problem1} is given by
$$ \tilde{X}(T) \ := \ \tilde{X}_{\lambda^*}(T) \ = \ H_{\lambda^*}(\tilde{Y}_{\lambda^*}(T)). $$
$\tilde{X}\tilde{Y}$ is a uniformly integrable martingale on
$[0,T]$. 
Furthermore, the functions $u$ and $w_{\lambda^*}$ defined
respectively in (\ref{utility max}) and (\ref{def primal value
function}) are different up to a constant in the way that
\begin{equation} \label{eq u = w + lambda}
u(x)\ = \ w_{\lambda^*}(x) +\lambda^* x_1.
\end{equation}
\item[(ii)] $u(x) < +\infty$ for all $x > 0$.
The function $u$ is increasing, continuously differentiable and strictly concave on $(0,+\infty)$. $u$ and $z_{\lambda^*}+\lambda^* x_1$ are conjugate, i.e., it holds that
\begin{eqnarray*}
z_{\lambda^*}(y) +\lambda^* x_1 & = & \sup_{x > 0} \, \{u(x) - xy\}, \quad y > 0; \\
u(x) & = & \inf_{y \geq 0} \, \{z_{\lambda^*}(y)+\lambda^* x_1 + xy\} , \quad x > 0.
\end{eqnarray*}
Moreover, $u$ satisfies
$$ u'(0) \ := \ \lim_{x \searrow 0} u'(x) \ = \ +\infty \qquad \text{and} \qquad u'(+\infty) \ := \ \lim_{x \to \infty} u'(x) \ = \ 0.$$
\item[(iii)] For $x>0$ it holds that
\begin{eqnarray*}
x u'(x) & = & \E \left[\tilde{X}(T) U'(\tilde{X}(T)) \right] + \lambda^* \E \left[\tilde{X}(T) L'(-\tilde{X}(T)) \right].
\end{eqnarray*}
\item[(iv)] It holds for the asymptotic elasticity of $u$ that
$$ AE(u)_{+} \ \leq \ AE(U - \lambda^{*} L)_{+} \ < \ 1. $$
\end{enumerate}
\end{theorem}

The proof of the theorem needs some auxiliary results which are stated first.

\begin{lemma} \label{lemma a,b}
Let Assumption \ref{ass nonempty q}, \eqref{ass u finite} and
\eqref{def W} hold true. Then
 for any $\lambda \geq 0$ we have the following results.
\begin{enumerate}
\item[(a)] $w_\lambda(x) < +\infty$ for all $x > 0$. There exits $y_0 > 0$ such that $z_\lambda(y) < +\infty$ for any $y > y_0$. The functions $w$ and $z$ are conjugate, i.e., it holds that
\begin{eqnarray*}
z_\lambda(y) & = & \sup_{x > 0} \{w_\lambda(x) - xy\}, \quad y > 0; \\
w_\lambda(x) & = & \inf_{y \geq 0} \{z_\lambda(y) + xy\}, \quad x > 0.
\end{eqnarray*}
The function $w_\lambda$ is increasing, continuously differentiable on $(0,+\infty)$ and the function $z_\lambda$ is strictly convex on $(y_0,+\infty)$. The functions $w_\lambda'$ and $z_\lambda'$ satisfy
$$ w_\lambda'(0) \ := \ \lim_{x \searrow 0} w_\lambda'(x) \ = \ +\infty \quad \mbox{ and } \quad z_\lambda'(+\infty) \ := \ \lim_{y \rightarrow \infty} z_\lambda'(y) \ = \ 0. $$
\item[(b)] If $z_\lambda(y)<+\infty$, then the optimal solution $\tilde{Y}_\lambda \in \mathcal{Y}(y)$ to problem exists and is unique.
\end{enumerate}
\end{lemma}

{\em Proof.}
By the property that $W_{\lambda}$ is a utility function for any $\lambda \geq 0$ (cf.\ Proposition \ref{prop properties w}) and by $w_{\lambda}(x) <+\infty$ for some $x$ (cf.\ Lemma \ref{lemma w finite}), the results follow from Theorem 2.1 in \cite{kramkov schachermayer}.
\hfill{$\Box$}

\begin{lemma} \label{lemma c-g}
Let Assumption \ref{ass nonempty q}, \eqref{ass u finite} and
\eqref{def W} hold true. Moreover, let $AE(W_{\lambda}) < 1$ for all
$\lambda \geq 0$. Then we have the following results.
\begin{enumerate}
\item[(a)] $z_\lambda(y) < +\infty$ for all $y > 0$. The functions $w_\lambda$ and $z_\lambda$ are continuously differentiable on $(0,+\infty)$ and the functions $w_\lambda'$ and $-z_\lambda'$ are strictly decreasing. They satisfy
$$ w_\lambda'(+\infty) \ := \ \lim_{x \rightarrow \infty} w_\lambda'(x) \ = \ 0 \quad \mbox{ and } \quad -z_\lambda'(0) \ := \ \lim_{y \rightarrow 0} -z_\lambda'(y) \ = \ +\infty. $$
\item[(b)] The optimal solution $\tilde{Y}_\lambda \in \mathcal{Y}(y)$ to problem \eqref{def dual value function} exists and is unique.
\item[(c)] The optimal solution $\tilde{X}_\lambda \in \mathcal{X}(x)$ to Problem \ref{problem1.1} exists and is unique.
If $\tilde{Y}_\lambda \in \mathcal{Y}(y)$ is the optimal solution to
problem \eqref{def dual value function} with $y=w_{\lambda}'(x)$,
then the dual relation yields to
$$ \tilde{X}_\lambda(T) \ = \ H_{\lambda}(\tilde{Y}_\lambda(T)) \quad \text{ and }
\quad \tilde{Y}_\lambda(T) \ = \ W_{\lambda}'(\tilde{X}_\lambda(T)).
$$ The process $\tilde{X}_\lambda \tilde{Y}_\lambda$ is a uniformly
integrable martingale over $[0,T]$.
\item[(d)] It holds for $w_{\lambda}'$ and
$z_{\lambda}'$ that
\begin{eqnarray*}
w_{\lambda}'(x) & = & \E \left[\frac{\tilde{X}_\lambda(T)
W_{\lambda}'(\tilde{X}_\lambda(T))}x \right] \quad \text{ and }
\quad z_{\lambda}'(y) \ = \ \E \left[ \frac{\tilde{Y}_\lambda(T)
Z_{\lambda}'(\tilde{Y}_\lambda(T))}y \right].
\end{eqnarray*}
\item[(e)] The value function $z_\lambda$ can be also expressed by
\begin{eqnarray} \label{eq different expression z}
z_\lambda(y) & = & \inf_{Q \in \mathcal{Q}} \, \E \left[Z_\lambda\left(y \frac{dQ}{dP}\right)\right],
\end{eqnarray}
where $dQ/dP$ denotes the Radon-Nikodym derivative of $Q$ with respect to $P$ on $(\Omega,\mathcal{F}_T)$.
\end{enumerate}
\end{lemma}

{\em Proof.}
It holds by Theorem 2.2 in \cite{kramkov schachermayer}.
\hfill{$\Box$}

\begin{lemma} \label{lemma existence lambda}
Let Assumption \ref{ass nonempty q}, \eqref{ass u finite} and
\eqref{def W} hold true. Moreover, let $AE(W_{\lambda}) < 1$ for all
$\lambda \geq 0$. Then there exists $\lambda^{*}\geq 0$ such that
$$ \E\left[L\left(-H_{\lambda^*}(\tilde{Y}_{\lambda^*}(T))\right)\right] \ = \ x_1. $$
\end{lemma}

{\em Proof.} 
First, let us assume that $\tilde{Y}_\lambda(T)/y=dQ/dP$, for some $Q \in \mathcal{Q}$. Then
it holds for any $\lambda \geq 0$ that $\tilde{X}_{\lambda}$ with
$\tilde{X}_{\lambda}(T)=H_{\lambda} \left(y \frac{dQ}{dP}\right)$ is
a uniformly integrable martingale under $Q$ (cf.\ \cite{kramkov schachermayer}, Theorem 2.2 (iii)), i.e.,
$$ x \ = \ \tilde{X}_{\lambda}(0) \ = \ \E_Q\left[\tilde{X}_{\lambda}(T)\right] \ = \ \E_Q\left[H_{\lambda} \left(y \frac{dQ}{dP}\right)\right]. $$
Therefore we have that $H_\lambda\left(y \frac{dQ}{dP}\right) \in
\mathcal{L}^1_T(\Omega,\mathcal{F},Q)$, and with the martingale
representation theorem it holds for any $t \in [0,T]$ that
$$ \tilde{X}_{\lambda}(t) \ = \ \tilde{X}_{\lambda}(0) + \int_0^t \pi'_u \, dS_u \ = \ x + \int_0^t \pi'_u \, dS_u. $$
Therefore, it holds that $\tilde{X}_{\lambda} \in \mathcal{X}(x)$. Moreover, it holds by \eqref{ass u finite} and the concavity of $U$ that
$$u(x) \ = \ \sup_{X \in \mathcal{A}(x)} \, \E[U(X(T))] \ < \ +\infty $$
for all $x > 0$, which implies that $U\left(H_\lambda\left(y \frac{dQ}{dP}\right)\right) \in \mathcal{L}^1_T(\Omega,\mathcal{F},P)$. \\
Finally, we show that $L\left(-H_\lambda\left(y
\frac{dQ}{dP}\right)\right) \in
\mathcal{L}^1_T(\Omega,\mathcal{F},P)$. Indeed, let us assume that
$$\E\left[L\left(-H_\lambda\left(y \frac{dQ}{dP}\right)\right)\right] \ = \ +\infty.$$ 
Then we have
\begin{eqnarray*}
\E\left[W_\lambda\left(H_\lambda\left(y \frac{dQ}{dP}\right)\right)\right] & = & \E\left[U\left(H_\lambda\left(y \frac{dQ}{dP}\right)\right) - \lambda L\left(-H_\lambda\left(y \frac{dQ}{dP}\right)\right)\right] \\
& = & \underbrace{\E\left[U\left(H_\lambda\left(y \frac{dQ}{dP}\right)\right)\right]}_{< +\infty} - \lambda \underbrace{\E\left[L\left(-H_\lambda\left(y \frac{dQ}{dP}\right)\right)\right]}_{=+\infty} \ = \ - \infty.
\end{eqnarray*}
But by (e), $\tilde{X}_{\lambda}(T)=H_\lambda\left(y \frac{dQ}{dP}\right)$ is the optimal solution
to $\sup_{X \in \mathcal{X}(x)} \E[W_\lambda(X(T))]$ -- a contradiction. 
Therefore, it holds that $\E\left[L\left(-H_\lambda\left(y \frac{dQ}{dP}\right)\right)\right] < +\infty$.
The existence of $\lambda^* > 0$ such that
$$ \E\left[L\left(-H_{\lambda^*} \left(y \frac{dQ}{dP}\right)\right)\right] \ = \ x_1, $$
was then shown by Lemma 6.1 of \cite{gundel weber}.
Now, let us assume that $\tilde{Y}_\lambda/y \in \mathcal{Y}(1)
\backslash D(\mathcal{Q})$. We follow the idea of \cite{kramkov
schachermayer}. Set $\hat{S} := (1,1/\tilde{X}_\lambda,S^1/\tilde{X}_\lambda, \ldots,S^m/\tilde{X}_\lambda)$
and since $\tilde{X}_\lambda\tilde{Y}_\lambda$ is a uniformly
integrable martingale, we can define $N_t := \tilde{X}_\lambda(t)\tilde{Y}_\lambda(t)/(xy)$
as a density process for probability measure $\tilde{Q}$, i.e.\ 
$N_T=d\tilde{Q}/dP$. Then $\tilde{Q}$ is an equivalent local
martingale measure for $\tilde{S}$, i.e.\ $\tilde{Q} \in
\mathcal{Q}(\tilde{S})$. Again, we can use the same arguments as
above.
\hfill{$\Box$} \\[-0.2cm]

Summarizing the statements above, we shall prove the main theorem. \\[-0.2cm]

{\em Proof of Theorem \ref{lemma incomplete market}.}
\begin{enumerate}
\item[(i)] By \eqref{def primal value function}, Lemma \ref{lemma c-g} (c) and  Lemma \ref{lemma existence lambda} it holds that
\begin{eqnarray} \label{eq w for optimal x}
w_{\lambda^*}(x) & = & \sup_{X \in \mathcal{X}(x)} \, \E [U(X_T) - \lambda^* L(-X_T)] \nonumber \\
& = & \E\left[U\left(H_{\lambda^*} \left( \tilde{Y}_{\lambda^{*}}(T) \right)\right)\right]
- \lambda^* \, \E \left[L\left(-H_{\lambda^*} \left( \tilde{Y}_{\lambda^{*}}(T) \right)\right)\right] \nonumber \\
& = & \E[U(X_{\lambda^*}(T))] - \lambda^* x_1.
\end{eqnarray}
For any $X \in \mathcal{A}(x)$, we have $\E[L(-X(T))] \leq x_1$ hence it holds that
\begin{eqnarray*}
\E[W_{\lambda^*}(X(T))] & = & \E[U(X(T)) - \lambda^*L(-X(T))] \ = \ \E[U(X(T)) - \lambda^*(L(-X(T)) - x_1)] - \lambda^* x_1 \nonumber \\
& = & \E[U(X(T))] - \lambda^*(\E[L(-X(T))] - x_1) - \lambda^* x_1 \ \geq \ \E[U(X(T))] - \lambda^* x_1,
\end{eqnarray*}
where the first identity follows from \eqref{def W}. Because
$\tilde{X}_{\lambda^*}$ is the unique wealth process that attains
the supremum in \eqref{eq w for optimal x}, we have the inequalities
\begin{eqnarray*}
\E\left[U(\tilde{X}_{\lambda^*}(T))\right] - \lambda^* x_1 = w_{\lambda^*}(x) = \E\left[W_{\lambda^*}(\tilde{X}_{\lambda^*}(T))\right] \geq \E[W_{\lambda^*}(X(T))] \geq \E[U(X(T))] - \lambda^* x_1.
\end{eqnarray*}
They become equalities if and only if $X=\tilde{X}_{\lambda^*}$,
which is in $\mathcal{A}(x)$ by Lemma \ref{lemma c-g} (c) and Lemma
\ref{lemma existence lambda}. Hence
$\tilde{X}=\tilde{X}_{\lambda^*}$ is the unique wealth process that
attains the supremum in
\begin{eqnarray*}
u(x) & = & \sup_{X \in \mathcal{A}(x)} \, \E[U(X(T))],
\end{eqnarray*}
which implies that $u(x) = \E[U(\tilde{X}(T))] = w_{\lambda^*}(x) + \lambda^* x_1$.
The uniform integrability also follows from Lemma \ref{lemma c-g}
(c).
\item[(ii)] The first results follow immediately from Lemma \ref{lemma a,b} (a) by putting $\lambda^{*}$ instead of $\lambda$.
By \eqref{eq u = w + lambda} which implies $u'(x)=w'_{\lambda^*}(x)$ for all $x \in (0,+\infty)$ it follows: $u'(0)=+\infty$.
\item[(iii)] This follows from \eqref{eq u = w + lambda} and Lemma \ref{lemma c-g} (d) where we write $\lambda^{*}$ instead of $\lambda$.
\item[(iv)] By the relation \eqref{eq u = w + lambda} and by the fact that $\lambda^{*} x_{1} \geq 0$ it holds that
\begin{eqnarray*}
AE(u)_{+} & = & \limsup_{x \to \infty} \frac{xu'(x)}{u(x)} \ = \ \limsup_{x \to \infty} \frac{xw_{\lambda^{*}}'(x)}{w_{\lambda^{*}}(x) + \lambda^{*}x_{1}} \\
& \leq & \limsup_{x \to \infty} \frac{xw_{\lambda^{*}}'(x)}{w_{\lambda^{*}}(x)} \ = \ AE(w_{\lambda^{*}})_{+} \ \leq \ AE(W_{\lambda^{*}})_{+} = AE(U - \lambda^{*} L)_{+} \ < \ 1,
\end{eqnarray*}
where the last two inequalities follow from Theorem 2.2 (i) of
\cite{kramkov schachermayer} and the assumption on
$AE(W_{\lambda})$.
\hfill{$\Box$}
\end{enumerate}

\begin{remark} {\em
Extending the results of Kramkov and Schachermayer in \cite{kramkov schachermayer2}, it holds that the assumption that the asymptotic elasticity of the function $W_{\lambda}$ is only sufficient. The necessary and sufficient condition for an optimal solution is that the value function of the dual problem is finite for all $y > 0$. In our model, the value function to the dual problem is
$$z_{\lambda^*} (y) \ = \ \inf_{Y \in \mathcal{Y}(y)} \ \E \left[Z_{\lambda^*}(Y_T)\right], $$
where $\lambda^* \geq 0$ is again such that $\E[L(-X(T))] = x_1$. It
follows from the definition of $W_{\lambda^*}$ in \eqref{def W} and
the fact that $W_{\lambda^*}$ has the properties of a utility
function (cf.\ Proposition \ref{prop properties w}).
} \end{remark}

\begin{lemma} \label{lemma incomplete case finite assumption}
The condition $z_\lambda(y) < +\infty$ for all $y > 0$ is equivalent to
$$\inf_{Q \in \mathcal{Q}} \, \E \left[Z_\lambda\left(y \frac{dQ}{dP}\right)\right] \ < \ +\infty, \quad \text{ for all }y > 0.$$
\end{lemma}

{\em Proof.}
The one direction follows immediately from property (g) in the proof of Theorem \ref{lemma incomplete market}. The other direction follows due to the property that the density processes $dQ/dP$ of equivalent martingale measures $Q$ belong to $\mathcal{Y}(1)$.
\hfill{$\Box$} \\[-0.2cm]

For solving Problem \ref{problem1}, the claim $AE(W_\lambda) < 1$ can therefore be replaced by $z_\lambda(y) < +\infty$. 
In the special case where the loss function $L$ is nonnegative, this holds true. 
The assertions of Theorem \ref{lemma incomplete market} are still valid which is stated as the next proposition.

\begin{proposition}
Let Assumption \ref{ass nonempty q}, \eqref{ass u finite} and \eqref{def W} hold true. 
Let furthermore the asymptotic elasticity of $U$ be strictly less than one and let the loss function $L$ be nonnegative-valued.
Then all the properties of Theorem \ref{lemma incomplete market} hold true.
\end{proposition}

{\em Proof.}
Let us suppose that $AE(U)<1$. By Note 2 in \cite{kramkov schachermayer2}, this implies
$$v(y) \ := \ \inf_{Q \in\mathcal{Q}}\E\left[V\left(y \frac{dQ}{dP}\right)\right] \ < \ +\infty$$
for all $y>0$. By Theorem \ref{lemma property Z} (ii), it holds for all $y \in (0,+\infty)$ that $Z_\lambda(y)  \leq  V(y)$ which consequently implies that
$$Z_\lambda\left(y \frac{dQ}{dP}\right) \ \leq \ V\left(y \frac{dQ}{dP}\right), \qquad \inf_{Q \in \mathcal{Q}} \, Z_\lambda\left(y \frac{dQ}{dP}\right) \ \leq \ \inf_{Q \in \mathcal{Q}} \, V\left(y\frac{dQ}{dP}\right),$$ 
and by equation \eqref{eq different expression z}: $z_\lambda(y)  \leq  v(y)$. \\
This means that $v(y)<+\infty$ implies $z_\lambda(y) < +\infty$ for
all $y>0$. Because by Proposition \ref{prop properties w},
$W_\lambda$ has the properties of a utility function and
$z_\lambda(y)$ is the value function of the dual problem to the
utility maximization problem $w_\lambda(x)= \sup_{\mathcal{X}(x)}
\E[W_\lambda(X_T)]$, we can apply Theorem 2 in \cite{kramkov
schachermayer2} to the $W_\lambda$ utility maximization problem to
derive Lemma \ref{lemma a,b} and Lemma \ref{lemma c-g}. Therefore,
the properties in Theorem \ref{lemma incomplete market} hold true.
\hfill{$\Box$}



\subsection{Optimal investment and consumption} \label{subsec consumption}

Because our approach is essentially developing the stochastic
version of the Legendre-Fenchel transform for solving convex
optimization problems, it can be extended to more complicated cases.
To illustrate this claim, this section considers the optimization
problem where a {\em cumulative consumption process} $C$ is added,
following the framework of \cite{karatzas zitkovic}.

Let us exactly define the process $C=(C_t)_{0 \leq t \leq T}$ as a nonnegative, nondecreasing, $\mathcal{F}$-adapted, RCLL process. 
We call the pair $(\pi,C)$ satisfying the above assumptions an {\em investment-consumption strategy}. 
The wealth process $X^{\pi,C,x}$ of the investor is given by
$$ X^{\pi,C,x}(t) \ = x + \int_0^t \pi_u' \, dS_u - C_t, \quad 0 \leq t \leq T. $$
The strategy $(\pi,C)$ is {\em admissible} if $X^{\pi,C,x}(T) \geq 0$. When there is no confusion, we simply write $X:=X^{\pi,C,x}$.
Furthermore, we call the consumption process $C$ {\em admissible} if there is a strategy $\pi$ such that $(\pi,C)$ is admissible. 
Suppose there is a probability measure $\mu$ such that
$$ C_t \ = \int_0^t c(u) \, \mu(du), \quad 0 \leq t \leq T, $$
where $c$ is the corresponding density processes. 
The set of all such density processes will be denoted by $\mathcal{A}^\mu(x)$. 
By this expression above, the terminal wealth is interpreted as the instantaneous consumption from time $T-$ to $T$ and is given by
\begin{eqnarray*}
X(T) & = & C_T - C_{T-} \ = \ \int_0^T c(u) \, \mu(du) - \int_0^{T-} c(u) \, \mu(du) \ = \ \frac12 c(T).
\end{eqnarray*}
Because the terminal wealth $X(T)$ can be expressed in terms of the consumption process $c$, it suffices to optimize over the consumption $c$ only. The optimal solution can be recovered from the optimal solution to the following pure consumption problem.
In this subsection we will focus on {reasonable elastic utility random fields}, which are utility functionals in time, wealth and random scenarios. For the exact notation and properties we refer to \cite{karatzas zitkovic}.

\begin{problem}\label{problem5}
Find an optimal consumption process $\tilde{c}$ and an optimal terminal wealth $X$ that achieve the maximum expected utility
\begin{equation} \label{eq utility maximization with consumption and risk constraint}
\begin{array} {c} u(x) \ = \ \sup \limits_{c \in \mathcal{A}^{\mu}(x)} \, \left\{ \E \left[ \int_0^T U_1(t, c(t))dt + U_2(c(T)/2) \right] \right\} \\ \\
\mbox{subject to } \ \E[L(-c(T)/2)] \ \leq \ x_1, \end{array}
\end{equation}
where $U_1$ is a deterministic utility random field with corresponding $K_{1}$ and $K_{2}$ (cf.\ Definition 3.1.\ in \cite{karatzas zitkovic}), $U_2$ a utility function and $L$ a loss function as defined in Definition \ref{def loss function} such that 
\begin{equation}\label{eq prop7.3}
0 \ < \ \liminf_{x \to
\infty} \frac{U'_2(x)}{K_1(x)} \ \leq \ \limsup_{x \to \infty}
\frac{U'_2(x)}{K_1(x)} \ < \ +\infty .
\end{equation}
\end{problem}

For the derivation of an optimal solution to Problem \ref{problem5}, we need the following assumption.

\begin{assumption} \label{assumption u finite}
There exists $x > 0$ such that $u(x)  < +\infty$. 
\end{assumption}

Same as in subsection \ref{sec incomplete}, we
reformulate the optimization problem under constraints by
introducing a Lagrange multiplier $\lambda \geq 0$. Defining
$W_\lambda(x):=U_2(x) - \lambda L(-x)$, we know by Proposition
\ref{prop properties w} that $W_\lambda$ is again a utility
function. 
Following Example 3.11.\ in \cite{karatzas zitkovic} for solving this optimization problem, the two utility measures $U_1$ and $W_{\lambda}$ by one utility random field $\mathcal{W}_\lambda:[0,T] \times \Real_+ \to \Real$ defined as
\begin{equation}\label{def W lambda}
\mathcal{W}_\lambda(t,x) \ := \ \left\{ \begin{array} {ll} 2TU_1(t,
\frac{x}{2T}), &  t < T; \\ 2W_\lambda(\frac{x}2), &  t=T.
\end{array} \right.
\end{equation}


We shall need the unconstrained optimization problem below to help
solve Problem \ref{problem5}.

\begin{problem}\label{problem4}
Find an optimal consumption process $\tilde{c}$ that achieves the maximum expected utility
\begin{equation*} 
w_\lambda(x) \ = \ \sup \limits_{c \in \mathcal{A}^{\mu}(x)} \, \E
\left[\int_0^T \mathcal{W}_\lambda(t,c(t)) \, \mu(dt) \right].
\end{equation*}
\end{problem}

The dual version of Problem \ref{problem4} is given by
\begin{equation} \label{eq dual problem optimal consumption}
z_\lambda(y) \ = \ \inf_{Q \in \mathcal{D}} \, \E \left[\int_0^T \sup_{x>0} \left(\mathcal{W}_\lambda(t,yY^Q_t) - xyY^Q_t\right) \, \mu(dt) \right] ,  
\end{equation}
where $\mathcal{D}$ denotes the domain of the dual problem, i.e., the closure of the set of all supermartingale measures of the stock
process $S$, and its elements are finitely-additive probability measures. 
The process $Y^Q$ is a supermartingale version for the density process of the maximal countably additive measure on $\mathcal{F}$ that is dominated by $Q$ (the regular part of $Q$, cf.\ \cite{karatzas zitkovic}).\\
Now, we present the main result of this subsection.

\begin{theorem} \label{theorem solution for optimal consumption}
Suppose Assumption \ref{assumption u finite} and \eqref{eq prop7.3}
hold true. Let $U_2$ and $L$ be such that $AE(U_2) < 1$ and
$W_\lambda(+\infty)> 0$. Then Problem \ref{problem5} has an optimal
solution $\tilde{c} \in \mathcal{A}^\mu(x)$ which is given by
$$ \tilde{c}(t) \ = \ \left\{ \begin{array} {ll} 2T(\partial_2U_1(t,\cdot))^{-1}\left(yY^{\tilde{Q}_t^y}\right), &  t < T; \\
2(W'_{\lambda^{*}})^{-1}\left(yY^{\tilde{Q}_T^y}\right), & t=T,
\end{array} \right.  $$
where $y=w_{\lambda^*}'(x)$ and $\tilde{Q}^y$ is a solution to the dual problem \eqref{eq dual problem optimal consumption}. \\
The corresponding optimal terminal wealth is given by
$$ \tilde{X}(T) \ = \ H_{\lambda^{*}}\left(yY^{\tilde{Q}_T^y}\right), $$
where $H_{\lambda^{*}}:=(W'_{\lambda^{*}})^{-1}$ denotes the inverse
of the first order derivative of $W_{\lambda}^{*}$ and $\lambda^{*}
\geq 0$ is such that $\E[L(-\tilde{X}(T))] = x_1$.\\
Moreover, the value functions $u$ and $z_{\lambda^*}$ have the bi-dual relation as in Theorem \ref{lemma incomplete market} (ii).
\end{theorem}

{\em Outline of Proof.} By the assumptions and by the properties of $L$ (cf.\ Definition \ref{def loss function}) as well as the properties of the asymptotic elasticity of $W_\lambda$ (cf.\ Lemma \ref{lemma ae of w}), it holds that $AE(W_\lambda) < 1$ and $0 < \liminf_{x \to \infty} \frac{W'_\lambda(x)}{K_1(x)} \leq \limsup_{x \to \infty} \frac{W'_\lambda(x)}{K_1(x)} < +\infty$. 
Therefore, by Example 3.11.\ in \cite{karatzas zitkovic}, $\mathcal{W}_\lambda$ is a reasonable elastic utility random field. \\
Now, using Theorem 3.10.\ (v) in \cite{karatzas zitkovic}, the optimal solution to Problem \ref{problem4} is given by
$$ \tilde{c}_\lambda(t) \ = \ \mathcal{I}_\lambda\left(t,yY^{\tilde{Q}_t^y}\right), \quad 0 \leq t \leq T, $$
where $\mathcal{I}_\lambda(t,y):=(\frac{d}{dx} \mathcal{W}_\lambda(t,x))^{-1}(y)$ is the inverse of the first order derivative of $\mathcal{W}_\lambda$, $y=w_\lambda'(x)$ and $\tilde{Q}^y$ is a solution to the dual problem \eqref{eq dual problem optimal consumption}.  
For $\mathcal{I}_\lambda$ it holds that
$$ \mathcal{I}_\lambda(t,y) \ = \ \left\{ \begin{array} {ll} 2T\left(\frac{d}{dx}U_1(t,x)\right)^{-1}(y), &  t < T; \\
2(W'_\lambda)^{-1}(y), &  t=T. \end{array} \right. $$ The optimal
terminal wealth of Problem \ref{problem4} is then given by
\begin{eqnarray*}
\tilde{X}_{\lambda}(T) & = & \frac12 \tilde{c}(T) \ = \
(W'_\lambda)^{-1}\left(yY^{\tilde{Q}_T^y}\right) \ = \
H_\lambda\left(yY^{\tilde{Q}_T^y}\right).
\end{eqnarray*}
Now, again choose $\lambda^{*} \geq 0$ such that
$\E[L(-\tilde{X}_{\lambda}(T))] = x_1$. Proof of the existence of
such a $\lambda^{*}$ is similar to that in Lemma \ref{lemma
existence lambda}. Similar to the proof of Theorem \ref{lemma
incomplete market} (i), we can show that
$\tilde{c}:=\tilde{c}_{\lambda^{*}}$
is the optimal consumption of Problem \ref{problem5}. Hence $\tilde{X}:=\tilde{X}_{\lambda^{*}}$ is the optimal terminal wealth.\\
The bi-dual relation between $u$ and $z_{\lambda^*}$  can be proved similar to that in Theorem \ref{lemma incomplete market} (ii), using the bi-dual relation $w_\lambda$ and $z_{\lambda}$ by Theorem 3.10.\ (iii) of \cite{karatzas zitkovic}.
\hfill{$\Box$}



\section{Solution in complete market}\label{sec complete market}

Let us now consider the case of a complete market, i.e., the set $\mathcal{Q}$ consists of only one element $Q$ -- the {\em unique} equivalent martingale measure. 
For $w_\lambda$ we can define the conjugate function $z_\lambda$ via
$$ z_\lambda(y) \ := \ \E \left[ Z_\lambda\left(y \frac{dQ}{dP}\right)\right]. $$

\subsection{Main theorem} \label{subsec main theorem complete market}

Again, our goal is now to solve the main optimization problem \ref{problem1}. 
In the complete market case, we do not need the assumption on the asymptotic elasticity of $W_\lambda$. 
The result is similar to Theorem \ref{lemma incomplete market}, except that it looks friendlier.

\begin{theorem} \label{lemma complete market}
Let Assumption \ref{ass nonempty q}, \eqref{ass u finite} and \eqref{def W} hold true.
Let $y_0:=\inf\{y > 0 \mid z(y) < +\infty\}$ and $x_0:=\lim_{y \searrow y_0} -z_\lambda'(y)$.
Then we have the following results.
\begin{enumerate}
\item[(i)] If $x < x_0$, then the optimal solution $\tilde{X} \in \mathcal{X}(x)$ to Problem \ref{problem1} is given by
$$ \tilde{X}(T) \ = \ H_{\lambda^*} \left(y \frac{dQ}{dP}\right), $$
for $y > y_0$, where it holds that $y=u'(x)$. 
$\lambda^{*} \geq 0$ is such that $\E[L(-\tilde{X}(T))] = x_1$.
$\tilde{X}$ is a uniformly integrable martingale under $Q$.
Furthermore, the functions $u$ and $w_{\lambda^*}$ defined respectively in (\ref{utility max}) and (\ref{def primal value function}) are different up to a constant in the way that
\begin{equation} \label{eq u gleich w plus lambda}
u(x)\ = \ w_{\lambda^*}(x) +\lambda^* x_1.
\end{equation}
\item[(ii)] $u(x) < +\infty$ for all $x > 0$.
The function $u$ is increasing, continuously differentiable on $(0,+\infty)$ and strictly concave on $(0,x_0)$. $u$ and $z_{\lambda^*}+\lambda^* x_1$ are conjugate, i.e.\ it holds that
\begin{equation*}
z_{\lambda^*}(y) +\lambda^* x_1 \ = \ \sup_{x > 0} \, \{u(x) - xy\}, \ y > 0; \quad
u(x) \ = \ \inf_{y \geq 0} \, \{z_{\lambda^*}(y)+\lambda^* x_1 + xy\} , \ x > 0.
\end{equation*}
\item[(iii)] For $0 < x < x_0$ it holds that $xu'(x) = \E[\tilde{X}(T)U'(\tilde{X}(T))] + \lambda^* \E[\tilde{X}(T) L'(\tilde{X}(T))]$. \\
Moreover, $u$ satisfies $ u'(0) := \lim_{x \searrow 0} u'(x) = +\infty$.
\end{enumerate}
\end{theorem}

{\em Proof.}
\begin{enumerate}
\item[(i)] First, since $W_{\lambda}$ is a utility function for any $\lambda \geq 0$ by
Proposition \ref{prop properties w}. By Theorem 2.0 (ii)  of
\cite{kramkov schachermayer}, the optimal solution to the problem in
(\ref{def primal value function}) is given by $\tilde{X}_{\lambda}(T) = H_{\lambda}(y dQ/dP)$, for  $x < x_0$ and $y > y_0$, where it holds that $y=w_{\lambda}'(x)$ or, equivalently, $x=-z_{\lambda}'(y)$.  \\
The existence of $\lambda^{*} \geq 0$, such that $\E[L(-H_{\lambda^*}(y dQ/dP))] = x_1$ was proven in Lemma \ref{lemma existence lambda}. \\
Furthermore, by \eqref{def primal value function} we have
\begin{eqnarray*} 
w_{\lambda^*}(x) & = & \sup_{X \in \mathcal{X}(x)} \, \E [U(X(T)) - \lambda^* L(-X(T))] \nonumber \\
& = & \E\left[U\left(H_{\lambda^*} \left(y \frac{dQ}{dP}\right)\right)\right] - \lambda^* \, \E \left[L\left(-H_{\lambda^*} \left(y \frac{dQ}{dP}\right)\right)\right] \ = \ \E\left[U(\tilde{X}_{\lambda^*}(T))\right] - \lambda^* x_1.
\end{eqnarray*}
Then we use the arguments in the proof of Theorem \ref{lemma incomplete market} (i). \\
Moreover, we have by \eqref{def H} and Theorem 2.0 (iii) of
\cite{kramkov schachermayer} that
$$ \E_Q [\tilde{X}(T)] \ = \ \E_P \left[ H_{\lambda^*} \left(y \frac{dQ}{dP}\right) \frac{dQ}{dP}\right] \ = \ \E_P \left[ -Z_{\lambda^*}' \left(y \frac{dQ}{dP}\right) \frac{dQ}{dP}\right] \ = \ -z_{\lambda^*}'(y) \ = \ x. $$
Therefore, $\tilde{X}$ is a $Q$-martingale and it belongs to
$\mathcal{X}(x)$.
\item[(ii)] It follows from \eqref{eq u gleich w plus lambda} and Lemma \ref{lemma a,b} (a).
\item[(iii)] The representation of $u'(x)$ follows from \eqref{eq u gleich w plus lambda} and the fact that $w_{\lambda}'(x) = \frac{1}{x}\ \E [\tilde{X}_{\lambda}(T) W_{\lambda}'(\tilde{X}_{\lambda}(T))]$, cf.\ Lemma \ref{lemma c-g} (d).
Moreover, this statement implies that $w_{\lambda^*}'(0) := \lim_{x \searrow 0} w_\lambda'(x) = +\infty$. 
\hfill{$\Box$}
\end{enumerate}

\subsection{Extensions in the Black-Scholes market} \label{sec extensions}

We assume now that we are within a Black-Scholes framework where the price processes are described by geometric Brownian motions.
Let $B=(B_1, \ldots, B_n)'$ be an $n$-dimensional Brownian motion on
$(\Omega, \mathcal{F}, (\mathcal{F}_t)_{0 \leq t \leq T}, P)$, where
the filtration $(\mathcal{F}_t)_{0 \leq t \leq T}$ is generated by
$B$. Let us assume that the market consists of one risk-free bond
$S^0$ with a deterministic interest rate $r:[0,T] \to \mathbb{R}$,
which is given by $S^0_t:=\exp\{\int_0^t r_s ds\}$, for $t \in
[0,T]$. Furthermore, there are $n$ stocks, whereas their discounted
price processes $S^i$, $i=1, \ldots,n$, are modeled as
\begin{equation} \label{eq price process}
\left\{ \begin{array} {rclrcll}
dS^i_t & = & S^i_t \left((\mu^i_t-r_t) \ dt +\sum_{j=1}^n {\sigma^{ij}_t \ dB^j_t} \right); \\
 S^i_0 & = & s_i, & i=1,\ldots,m.
\end{array} \right.
\end{equation}

In the following the subscript $t$ is neglected. Here, $\mu^i$ and
$\sigma^{ij}$ are progressively measurable stochastic processes with
respect to the filtration $(\mathcal{F}_t)_{0 \leq t \leq T}$.
$\mu^i$ describes the drift of the $i$-th stock and
$\{\sigma^{ij}\}_{j=1}^n$ the volatilities of the $i$-th stock. Let
us define the volatility matrix $\sigma_t:=(\sigma^{ij}_t)_{n \times
n}$ and the risk premium process $\alpha=(\alpha^1, \ldots,
\alpha^n)'$ by $\alpha^i_t:=\mu^i_t - r_t$. We assume that $\alpha$
is uniformly bounded, $\sigma$ has full rank and that $\sigma
\sigma'$ is invertible and bounded. In this setting our market is
complete, because the number of assets is equal to the dimension of the
Brownian motion. Therefore, there exists a unique equivalent
martingale measure $Q$ and the Radon-Nikodym density $N$ is given by
\begin{equation} \label{eq definition rnd}
N_t \ := \ \left. \frac{dQ}{dP}\right|_{\mathcal{F}_t} \ = \ \exp \left\{ -\int_0^t {\theta'_s} \, dB_s - \frac{1}{2} \int_0^t {\left(||\theta_s||^2 \right)} \, ds \right\},
\end{equation}
where $\theta_t := \sigma_t^{-1} \alpha_t$ is the market price of risk. \\

For an initial capital $x > 0$, the wealth process is given by
\eqref{eq wealth process}. Using the price process dynamics
\eqref{eq price process}, we obtain the stochastic differential
equation
\begin{eqnarray} \label{eq dynamics wealth process}
\left\{ \begin{array} {rclrcll}
dX^{\pi,x}(t) & = & \pi'_t \diag(S_t) \alpha_t \, dt + \pi'_t \diag(S_t) \sigma_t \, dB_t; \\
 X^{\pi,0}(t) & = & x.
\end{array} \right. 
\end{eqnarray}

In a complete market, it is known that any admissible contingent
claim $\xi$ can be uniquely hedged. The wealth process of its
self-financing hedging portfolio $(x,\pi)$ evolves according to
\eqref{eq dynamics wealth process} and has time-$T$ payoff $\xi$.
Our optimal trading strategy $\pi^*$ for Problem \ref{problem1} is
unique, because it is part of the unique hedging portfolio
$(x,\pi^*)$ of a contingent claim whose payoff is the optimal
terminal wealth
$$X^{\pi^*,x}(T) \ = \ \tilde{X}(T) \ = \ H_{\lambda^*} \left(y \frac{dQ}{dP}\right)  = \ H_{\lambda^*} \left( y \exp \left\{ -\int_0^T {\theta'_s}  dB_s - \frac{1}{2} \int_0^T {\left(||\theta_s||^2 \right)} ds \right\} \right) $$
by Theorem \ref{lemma complete market} (i), where
$y=w_{\lambda^*}'(x)$ and $\lambda^*$ is such that
$\E[L(-X^{\pi^*,x}(T))]=x_1$.\\

An explicit form of the optimal portfolio is only possible when the market coefficients $\alpha$ and $\sigma$ are deterministic, cf.\ \cite{gabih}. 
The distribution of $ X^{{\pi^*},x}(T)$ is given by
$$ P\left( X^{{\pi^*},x}(T) \leq a \right) \ = \ \Phi \left(\frac{\ln(W'_{\lambda^{*}}(a)/y) + \frac12 \int_{0}^{T} ||\theta_{t}||^{2}dt}{\sqrt{\int_{0}^{T} ||\theta_{t}||^{2}dt}} \right), $$
where $\Phi$ denotes the distribution function of the standard normal distribution. We shall give an example for it in the next subsection.

\subsection{Example} \label{sec examples}

{\em Let $U(k)=- \frac1k + 1$ and $L(k)= - \frac3k$ be given. Then all
properties of Definitions \ref{def utility function} and \ref{def
loss function} are satisfied. Then we have $W_\lambda(k)=-\frac1k +
1 - \frac{3\lambda}k$ and
$H_\lambda(k)=\sqrt{\frac{1+3\lambda}{k}}$. Let us assume that
$\theta < \zeta$ are such that $r_{\min} \leq x_1 \leq r_{\max}$ for
$\theta < N_T < \zeta$. Then the optimal wealth process is given by
$$ \tilde{X}(t) \ = \ X^{{\pi^*},x}(t) \ = \ \frac1{N_t} \sqrt{\frac{1+3\lambda}{y}} \cdot \E\left[\left. N_t^{\frac12} \cdot (\exp(a+b\eta))^{\frac12} {\bf 1}_{\{\theta < N_t e^{a+b\eta} < \zeta\}} \, \right| \, \mathcal{F}_t \right], $$
where $a:=- \frac{1}{2} \int_t^T {\left(||\theta_s||^2 \right)} ds$,
$b:=-||\theta||$ and $\eta$ is a standard Gaussian random variable
independent of $\mathcal{F}_t$. Moreover, $\lambda^*$ is the unique
solution of $\left[e^{-\gamma \tilde{X}(T)}\right] = x_1$ and $y \in
(0,+\infty)$ is such that $\mathbb{E}[N_T \tilde{X}(T)]=x$. The
corresponding trading strategy is given by
\begin{eqnarray} \label{eq ex optimal trading}
\pi^*_t & = & -\diag(S_t)^{-1} (\sigma'_t)^{-1} \theta_t e^{\frac12a + \frac{b^2}4}
\sqrt{\frac{1+3\lambda}{y}N_t}  \cdot\left( -\frac1{2N_t}
\left(\Phi\left(\frac{\ln(\zeta/N_t) - a}b-\frac{b}2\right) \right. \right.  \\
& & \left. \left. - \Phi\left(\frac{\ln(\theta/N_t) -
a}b-\frac{b}2\right)\right) + \varphi\left(\frac{\ln(\zeta/N_t) -
a}b-\frac{b}2\right) \frac{1}{N_t\zeta b} -
\varphi\left(\frac{\ln(\theta/N_t) - a}b-\frac{b}2\right)
\frac{1}{N_t\theta b} \right), \nonumber
\end{eqnarray}
where $\varphi$ is the density of the cumulative standard-normal
distribution function $\Phi$.} \\

{\em Proof.}
The density $N_t$ of the equivalent martingale measure can be expressed by \eqref{eq definition rnd}, so it holds that
\begin{eqnarray*}
N_T & = & \exp \left\{ -\int_0^T {\theta'_s} \, dB_s - \frac{1}{2} \int_0^T {\left(||\theta_s||^2 \right)} \, ds \right\} \ = \ N_t \cdot \exp \left\{ -\int_t^T {\theta'_s} \, dB_s - \frac{1}{2} \int_t^T {\left(||\theta_s||^2 \right)} \, ds \right\} \\
& = & N_t \cdot \exp(a + b \eta),
\end{eqnarray*}
where $a:=- \frac{1}{2} \int_t^T {\left(||\theta_s||^2 \right)} ds$,
$b:=-||\theta||$ and $\eta$ is a standard Gaussian random variable
independent of $\mathcal{F}_t$. The process $N\tilde{X}$ is a
martingale with respect to $P$, so we have
\begin{eqnarray*}
N_t \tilde{X}_t & = & \E[ N_T X_T \, | \, \mathcal{F}_t] \\
\Leftrightarrow \ \tilde{X}_t & = & \E \left[ \left. \frac{N_T}{N_t} H_\lambda(yN_T) {\bf 1}_{\{\theta < N_T < \zeta\}} \, \right| \, \mathcal{F}_t \right] \ = \ \E \left[ \left. \frac{N_T}{N_t}
\sqrt{\frac{1+3\lambda}{yN_T}} {\bf 1}_{\{\theta < N_T < \zeta\}} \,
\right| \, \mathcal{F}_t \right].
\end{eqnarray*}
Following \cite{gabih} we can use the representation
$$ \frac{c}{N_t} \E[g(N_t,\eta) \, | \, \mathcal{F}_t] \ = \ \frac{c}{N_t} \psi(N_t) $$
with $\psi(z) = \E[g(z,\eta)]$ for $z \in (0,+\infty)$, where $g$ is
a measurable function and $c \in \Real$ is a constant, and derive
the process $X$ in the way that
\begin{eqnarray*}
\tilde{X}_t & = & \frac1{N_t} \sqrt{\frac{1+3\lambda}{y}} \cdot
\E\left[\left. N_t^{\frac12} \cdot (\exp(a+b\eta))^{\frac12} {\bf
1}_{\{\theta < N_t e^{a+b\eta} < \zeta\}} \, \right| \,
\mathcal{F}_t \right].
\end{eqnarray*}
Choose $g(z,x) = z^{\frac12} e^{\frac12(a+b\eta)} {\bf 1}_{\{\theta < z e^{a+bx} < \zeta\}}$ and with it we compute
\begin{eqnarray*}
\psi(z) & = & \E[g(z,\eta)] \ = \ \frac1{\sqrt{2\pi}} \int_{-\infty}^{+\infty} z^{\frac12} e^{\frac12(a+bx)} e^{-\frac12x^2} {\bf 1}_{\{\theta < ze^{a+bx} < \zeta\}} \, dx \\
& = & \frac{z^{\frac12}e^{\frac12a-\frac{b^2}4}}{\sqrt{2\pi}} \int_{\frac{\ln(\theta/z) - a}b}^{\frac{\ln(\zeta/z) - a}b}  e^{-\frac12(x-b/2)^2} \, dx \ = \ \frac{z^{\frac12}e^{\frac12a-\frac{b^2}4}}{\sqrt{2\pi}} \int_{\frac{\ln(\theta/z) - a}b-\frac{b}2}^{\frac{\ln(\zeta/z) - a}b-\frac{b}2}  e^{-\frac12x^2} \, dx \\
& = & z^{\frac12}e^{\frac12a-\frac{b^2}4} \left[\Phi\left(\frac{\ln(\zeta/z) - a}b-\frac{b}2\right)- \Phi\left(\frac{\ln(\theta/z) - a}b-\frac{b}2\right)\right].
\end{eqnarray*}
Now, set $\tilde{X}_t = \frac1{N_t} \sqrt{\frac{1+3\lambda}{y}}
\psi(N_t) = F(N_t,t)$ with
$$F(z,t) \ := \ z^{-\frac12} e^{\frac12a-\frac{b^2}4} \sqrt{\frac{1+3\lambda}{y}}  \left[ \Phi\left(\frac{\ln(\zeta/z) - a}b-\frac{b}2\right) - \Phi\left(\frac{\ln(\theta/z) - a}b-\frac{b}2\right)\right],  $$
it holds by It\^{o}'s formula that
\begin{eqnarray} \label{eq dynamics wealth process with f}
d\tilde{X}_t & = & F_t(N_t,t) \, dt + F_z(N_t,t) \, dN_t + \frac12 F_{zz}(N_t,t) \, dN_tdN_t \nonumber \\
& = & \left(F_t(N_t,t) + \frac12 F_{zz}(N_t,t) N_t^2 ||\theta_t||^2\right) \, dt - F_z(N_t,t) N_t \theta'_t \, dB_t,
\end{eqnarray}
where $F_z$, $F_{zz}$ and $F_t$ denote the partial derivatives of
$F(z,t)$ with respect to $z$ and $t$. Comparing the coefficients in
front of $dB_t$ in \eqref{eq dynamics wealth process} and \eqref{eq
dynamics wealth process with f}, we have that
\begin{equation*}
(\pi^*_t)' \diag(S_t) \sigma_t \ = \ - F_z(N_t,t) N_t \theta'_t \quad \Leftrightarrow \quad \pi^*_t \ = \ -\diag(S_t)^{-1} (\sigma'_t)^{-1} \theta_t N_t F_z(N_t,t).
\end{equation*}
Let us compute the first order derivative of $F(z,t)$ with respect
to $z$.
\begin{eqnarray*}
F_z(z,t) & = & e^{\frac12a + \frac{b^2}4} \sqrt{\frac{1+3\lambda}{y}}  \cdot\left( -\frac12 z^{-\frac32} \left[\Phi\left(\frac{\ln(\zeta/z) - a}b-\frac{b}2\right) - \Phi\left(\frac{\ln(\theta/z) - a}b-\frac{b}2\right)\right] \right. \\
& & \left. + z^{-\frac12} \left( \varphi\left(\frac{\ln(\zeta/z) - a}b-\frac{b}2\right) \frac{1}{z\zeta b} - \varphi\left(\frac{\ln(\theta/z) - a}b-\frac{b}2\right) \frac{1}{z\theta b} \right) \right),
\end{eqnarray*}
where $\varphi$ denotes the density function of the standard-normal distribution.
With this we get the expression \eqref{eq ex optimal trading}.
\hfill{$\Box$}

\section{Conclusion} \label{sec conclusion}

In this paper we solved the expected utility maximization problem under a utility-based shortfall constraint in a general incomplete market which admits no arbitrage. 
The utility function and the loss function therein do not need to have a special form. 
We only assumed that that the value function of the primal problem is real-valued, that the asymptotic elasticity of the utility function is smaller than one and that the loss function is non-negative. 
Moreover,we solved the problem in an optimal investment and consumption framework. 
In all cases the optimal terminal wealth has the same form as derived in Theorem \ref{lemma incomplete market} (i), i.e., the inverse of the first order derivative of the utility function combined with the loss function and a Lagrangian multiplier. \\[-0.2cm]

One interesting question for further research is whether the results can be extended to dynamic risk measures. Cuoco et al.\ \cite{cuoco et al} considered the problem under semi-dynamic risk constraints and derived solutions in a complete Black-Scholes market. Moreover, we will transfer the model in a setting of incomplete information, i.e., the investor has a partial knowledge about the market, described by a filtration $\mathbb{G}:=(\mathcal{G}_t)_{t \in [0,T]} \subset \mathbb{F}$:
$$ u_t(x) \ = \ \sup_{X \in \mathcal{X}_t(x)} \mathbb{E}[U(X(T)) \, | \, \mathcal{G}_t], \qquad \text{subject to} \quad \mathbb{E}[L(-X(T)) \, | \, \mathcal{G}_t] \leq x_1, \ P\text{-a.s.} $$
By considering the martingale representation under the filtration $\mathbb{G}$ for the wealth process may give the optimal solution for this problem. Dealing with incomplete information is one of our future research interests.


\end{document}